\documentclass[twocolumn]{aastex631}

\usepackage{soul}
\usepackage{bm}
\usepackage{amsmath}
\usepackage{tikz}
\usetikzlibrary{shapes.geometric, arrows, positioning, fit, backgrounds}
\graphicspath{{fig/}{./}}
\begin{document}

\newcommand{\fhc}[1]{{\bf{\color{blue}{#1}}}}
\newcommand{\wjc}[1]{{\color{red}{#1}}}
\newcommand{\mbh}{\ifmmode {M_{\bullet}} \else $M_{\bullet}$\fi}
\newcommand{\mvp}{\ifmmode {M_{\rm VP}} \else $M_{\rm VP}$\fi}
\newcommand{\rblr}{\ifmmode {R_{\rm BLR}} \else $R_{\rm BLR}$\fi}
\newcommand{\kms}{$\rm km\,s^{-1}$}
\newcommand{\feii}{\ensuremath{\mathrm{Fe~\textsc{ii}}}}
\newcommand{\hb}{\ensuremath{\mathrm{H}\beta}}
\newcommand{\ha}{\ensuremath{\mathrm{H}\alpha}}
\newcommand{\PaperI}{\hyperlink{cite.Li2024}{Paper I}}

\title{Velocity-resolved Ionization Mapping of the Broad-line Region. II. Signatures of Eccentric Broad-line Regions from a Quantitative Photoionization-Based Model}

\correspondingauthor{Hai-Cheng Feng}
\email{hcfeng@ynao.ac.cn}
\correspondingauthor{Xinwu Cao}
\email{xwcao@zju.edu.cn}

\author[0000-0002-2581-8154]{Jiancheng Wu}
\affiliation{Institute for Astronomy, School of Physics, Zhejiang University, 866 Yuhangtang Road, Hangzhou 310058, People's Republic of China}

\author[0000-0002-1530-2680]{Hai-Cheng Feng}
\affiliation{Yunnan Observatories, Chinese Academy of Sciences, Kunming 650216, Yunnan, People's Republic of China}

\author[0000-0003-3823-3419]{Sha-Sha Li}
\affiliation{Yunnan Observatories, Chinese Academy of Sciences, Kunming 650216, Yunnan, People's Republic of China}

\author[0000-0003-4773-4987]{Qingwen Wu}
\affiliation{Department of Astronomy, School of Physics, Huazhong University of Science and Technology, Luoyu Road 1037, Wuhan, China}

\author[0000-0002-2355-3498]{Xinwu Cao}
\affiliation{Institute for Astronomy, School of Physics, Zhejiang University, 866 Yuhangtang Road, Hangzhou 310058, People's Republic of China}
\affiliation{Center for Cosmology and Computational Astrophysics, School of Physics, Zhejiang University, 866 Yuhangtang Road, Hangzhou, 310058, People's  Republic of China}

\begin{abstract}
Velocity-resolved reverberation mapping (RM) is one of the most powerful tools for probing the geometry and kinematics of the broad-line region (BLR) in active galactic nuclei (AGNs). However, its interpretation may be non-unique, as similar velocity-delay signatures can arise from different combinations of BLR geometry and kinematics. To address this degeneracy, this series of studies develops velocity-resolved ionization mapping (IM) as a complementary technique. In this second paper of the series, we construct photoionization-based BLR models to quantitatively demonstrate the diagnostic power of the joint RM+IM framework. We calculate local \ha\ and \hb\ emissivities with \textsc{Cloudy}, couple them with disk-like BLR models, and generate synthetic velocity-dependent line fluxes, lags, and Balmer decrements (BDs). The model reproduces the observed anti-correlation between the integrated BD and the incident ionizing continuum, with scatter naturally introduced by reverberation delays. We show that a virialized elliptical BLR can produce diverse velocity-delay maps and velocity-resolved lags that mimic RM signatures commonly attributed to virialized motion, inflow, or outflow. In contrast, the velocity-resolved BD is sensitive to the symmetry of the BLR gas distribution, providing a key diagnostic for distinguishing eccentric geometry from non-virial motions. Through proof-of-concept comparisons with one double-peaked and one single-peaked AGN, we show that an elliptical-disk BLR can simultaneously reproduce the observed line profiles, velocity-resolved lags, and velocity-resolved BDs. These results provide a quantitative basis for IM, support the presence of eccentric BLR structures in AGNs, and demonstrate the potential of joint RM+IM analyses to break the BLR geometry-kinematics degeneracy. This framework has important implications for identifying geometry-induced biases and improving reverberation-based black hole mass measurements.
\end{abstract}


\section{Introduction} \label{sec:intro}
Over the past four decades, reverberation mapping \citep[RM;][]{Blandford1982, Peterson1993} has been one of the most powerful techniques for measuring the size, geometry, and kinematics of the broad-line region (BLR) in active galactic nuclei (AGNs). The method relies on a simple physical picture: gas in the BLR is photoionized by ultraviolet and X-ray radiation from the central ionizing source, and the resulting line emission responds to continuum variations after a light-travel-time delay. Therefore, the measured lag, $\tau$, between the driving-continuum and emission-line light curves can provide a characteristic BLR size, $\rblr \sim c\tau$, where $c$ is the speed of light. Because gas at different locations within the BLR is expected to exhibit different delays and line-of-sight velocities, analyzing how the delay varies as a function of velocity across the broad-line profile---the so-called velocity-resolved RM technique---allows one to further probe the geometry and kinematics of the BLR \citep[e.g.,][]{Bentz2008, Denney2010, Barth2011, Li2022a, Li2026, Feng2021a, Feng2024, Feng2025, Wang2025}.

Conventional RM interpretations often implicitly assume an axisymmetric BLR with azimuthally homogeneous properties at a given radius. In this framework, the diversity of velocity-resolved lag profiles is attributed primarily to different BLR kinematics \citep[e.g.,][]{Denney2009, Bentz2010, Du2016, Li2021, Yao2024}. For most AGNs, the BLR lies at distances of thousands of gravitational radii from the central supermassive black hole (SMBH), where relativistic effects are generally weak. Such a configuration can produce an approximately symmetric, often single-peaked broad-line profile. Observationally, however, many AGNs exhibit complex broad-line profiles, some of which evolve rapidly, challenging this simple picture of the BLR \citep[e.g.,][]{Eracleous2003, Schimoia2015, Storchi-Bergmann2017, Ilic2023, Marsango2024, Nagoshi2024, Wu2025a}. To account for these phenomena, several asymmetric BLR models have been proposed, for example involving an eccentric BLR geometry \citep{Eracleous1995} or a spiral-arm-like substructure \citep{Gilbert1999}. Such complexity in the BLR gas distribution can reshape the velocity-resolved lag profile independently of the underlying kinematics, thereby introducing degeneracies into conventional RM interpretations. Consequently, even a two-dimensional velocity-delay map may not yield a unique physical interpretation \citep[e.g.,][]{Pancoast2014, Pancoast2018, Williams2020, Horne2021}.

To overcome this limitation, in our previous work \citep[][; hereafter \PaperI]{Li2024}, we  proposed an ionization mapping (IM) method that infers the BLR structure by mapping the spatial distribution of the ionization properties of the gas. The basic principle is that the ionization state of BLR gas is governed jointly by the distance from the central ionizing source \citep{Korista2004}, which determines the incident ionizing photon flux, and by intrinsic gas properties such as density. In this sense, an ionization map of the BLR can serve as a promising tool for probing its geometry and even its detailed structure.

Accurate reconstruction of such a map requires combining photoionization calculations with dynamical modeling of the BLR. In practice, gas ionization states are commonly traced through emission-line ratios, among which the Balmer decrement (BD), usually defined as $\ha/\hb$, is one of the most widely adopted diagnostics. Although the BD alone cannot pinpoint the location of the emitting gas in the absence of detailed information about the BLR, variations in the BD at a given radius mainly reflect variations in the local gas distribution, because the incident continuum flux is nearly the same. As a result, velocity-resolved BDs provide an avenue for diagnosing asymmetries in the BLR. Combining RM with IM therefore offers a way to break the degeneracy between BLR geometry and kinematics.

In \PaperI, we applied the IM method to a sample of AGNs with simultaneous spectroscopic monitoring observations of $\ha$ and $\hb$. By jointly analyzing the velocity-resolved lags and BDs, we found that a longer lag is typically accompanied by a larger BD in most objects, supporting the basic expectation of IM. Intriguingly, in objects with asymmetric velocity-resolved lag profiles, the BDs show similar velocity-dependent patterns. This behavior poses a challenge to non-virial motion scenarios, such as inflow or outflow, that are commonly adopted in conventional RM studies, and qualitatively favors an eccentric BLR geometry. In this work, we further investigate this interpretation quantitatively by constructing photoionization-based BLR models with different geometric configurations and examining whether they can simultaneously reproduce the observed line profiles, velocity-resolved lags, and velocity-resolved BDs.

This work is structured as follows. Section~\ref{sec:2} describes the model. Section~\ref{sec:3} presents the simulated results and compares predictions with observations. Section~\ref{sec:4} explores implications for mass measurements, and Section~\ref{sec:5} summarizes our conclusions.

\section{Methodology}\label{sec:2}
The goal of this work is to quantitatively examine how different geometric and kinematic configurations of the BLR affect the velocity-dependent behaviors of the lags and BDs, as well as the broad-line profiles. This examination enables us to assess whether the observational features reported in \PaperI\ can be theoretically reproduced by plausible BLR configurations. For this purpose, we construct photoionization-based BLR models. In these models, photoionization calculations determine the local $\ha$ and $\hb$ emissivities, which are then coupled with the spatial distribution and velocity field of the BLR gas to generate synthetic velocity-resolved lags, velocity-resolved BDs, and line profiles. The details of the photoionization calculations, the geometric and kinematic configurations of the BLR, and the construction of synthetic observational signatures are described in the following subsections.

\subsection{Photoionization Calculations}\label{sec:2.1}
There are various approaches to modeling the emissivities of hydrogen recombination lines in photoionized gas. Among them, Case B recombination is one of the most widely used prescriptions, as it provides a simple and successful description for low-density nebular gas \citep{Osterbrock2006}. However, the BDs measured in BLRs are often found to deviate significantly from the canonical Case B expectation \citep{LaMura2007, Dong2008}. This discrepancy highlights the limited applicability of the Case B approximation to BLR gas, whose physical conditions differ substantially from those in classical nebulae \citep{Netzer1975}, including higher gas densities, larger optical depths in the emission lines, and a stronger ionizing radiation field. Moreover, the BD is sensitive to the local ionization conditions of the emitting gas \citep{Korista2004, Wu2023}, further suggesting that a simple Case B prescription is insufficient to describe spatially extended BLR gas spanning a broad range of ionization states. Motivated by these considerations, we use the photoionization code \textsc{Cloudy} \citep[v23.01;][]{Chatzikos2023} to compute the local \ha\ and \hb\ emissivities under physical conditions appropriate for the BLR.

We model the BLR gas as plane-parallel slabs of constant hydrogen density $n_{\rm H}$, illuminated by an incident AGN ionizing continuum. The strength of the radiation field is parameterized by the hydrogen-ionizing photon flux $\Phi({\rm H})$, which is related to the hydrogen-ionizing photon rate $Q({\rm H})$ (in units of s$^{-1}$) by
\begin{equation}\label{eq:1}
    \Phi({\rm H}) = \frac{Q({\rm H})}{4\,\pi\,r^2},
\end{equation}
where $r$ is the distance from the central source. For a given set of $n_{\rm H}$, $\Phi({\rm H})$, and total hydrogen column density $N_{\rm H}$, \textsc{Cloudy} solves for the thermal and ionization structure of the gas and returns the corresponding \ha\ and \hb\ emissivities. Since our goal is not to model a particular AGN, but rather to characterize the generic radial dependence of the Balmer-line emissivities in a typical BLR environment, we adopt a representative hydrogen number density of $n_{\rm H} = 10^{10}\,\mathrm{cm}^{-3}$ and a total hydrogen column density of $N_{\rm H} = 10^{23} \rm \, cm^{-2}$ \citep{Baldwin1995}. The resulting emissivities are stored as a lookup table and interpolated when evaluating gas at different radii.

The incident spectral energy distribution (SED) is constructed by combining a standard thin disk \citep[SSD;][]{Shakura1973} component in the UV--optical band with an empirical X-ray component. The SSD emission is modeled as a multi-temperature blackbody spectrum specified by the black hole mass $M_\bullet$ and dimensionless accretion rate $\dot{m}$. For the X-ray component, we follow the prescription adopted by \citet{Wu2024}. The 2~keV monochromatic luminosity $L_{\nu}(2\,{\rm keV})$ is determined from the optical-to-X-ray spectral index $\alpha_{\rm OX}$, which is related to the monochromatic luminosity at 2500~\AA{} through the observed correlation \citep{Steffen2006},
\begin{equation}\label{eq:2}
    \alpha_{\rm OX} = -0.14\,\log L_{\nu}(2500\,\text{\AA}) + 2.64,
\end{equation}
where $\alpha_{\rm OX} \equiv -0.3838\,\log(F_{2500\text{\AA}}/F_{2\,\text{keV}})$ \citep{Tananbaum1979}. The X-ray spectrum is modeled as a power law $L_\nu \propto \nu^{-(\Gamma-1)}$ anchored at 2~keV. The photon index $\Gamma$ is set by the Eddington ratio $\lambda_{\rm Edd}$ according to the empirical relation of \citet{Brightman2013}:
\begin{equation}\label{eq:3}
    \Gamma_{2\text{--}10\,\text{keV}} = 0.32\,\log\lambda_{\rm Edd} + 2.27.
\end{equation}
The final incident SED is obtained by combining the SSD spectrum at low frequencies with the X-ray power law at high frequencies. We then compute the total hydrogen-ionizing photon rate by integrating the composite SED from 13.6~eV to the upper boundary of the frequency grid \citep{Wu2023, Wu2024}.

The intrinsic BD at each BLR location is obtained directly from the tabulated emissivities by linear interpolation in log-space,
\begin{equation}\label{eq:4}
\mathrm{BD}=\epsilon_{\mathrm{H}\alpha}/\epsilon_{\mathrm{H}\beta},
\end{equation}
as a function of $\Phi({\rm H})$, or equivalently, as a function of radius for a given $Q({\rm H})$.

\subsection{BLR Geometry and Kinematics}\label{sec:2.2}
We model the BLR using the elliptical disk proposed by \citet{Eracleous1995}, and perform the numerical calculation following the implementation of \citet{Wu2026}, with minor modifications. In this framework, the line-emitting gas is distributed in a flattened disk-like geometry, and the gas clouds move on Keplerian orbits governed by the gravitational potential of the central SMBH. All orbits are assumed to share the same eccentricity and the same orientation of the major axis. This configuration allows the BLR to maintain a long-lived eccentric geometry while preserving virialized motion. An additional advantage of the elliptical disk model is that it can also recover an axisymmetric circular disk when the eccentricity is set to zero, making it flexible for comparison with traditional reverberation-mapping treatments.

The geometry of the model is determined by five parameters: the eccentricity $e$, the orientation of the major axis $\phi_0$, the inclination angle $i$, the inner pericenter distance $r_{\rm pi}$, and the outer pericenter distance $r_{\rm po}$. For a virialized BLR, each gas cloud follows an elliptical Keplerian orbit with eccentricity $e$ and pericenter distance $r_{\rm p}$. Its radial distance $r$ from the central SMBH is given by
\begin{equation}\label{eq:5}
    r(\phi) = \frac{r_{\rm p}(1 + e)}{1 - e\,\cos(\phi - \phi_0)},
\end{equation}
where $\phi$ is the azimuthal angle of the cloud in the disk plane. Assuming the observer is located at $\phi=0$, the projected line-of-sight velocity $v_{\rm los}$ is computed from the Keplerian velocity according to the adopted inclination $i$ and orbital phase $\phi - \phi_0$. The calculation is performed using the four-velocity formalism of \citet{Chen1989}, which accounts for relativistic effects. In addition to the orbital motion, the model also includes a Gaussian local broadening parameter, $\sigma_{\rm loc}$, to account for unresolved local velocity components, which may be associated with thermal or micro-turbulent motions. The local broadening inferred from the spectral fits of \citet{Eracleous1995} ranges from 800 to 1400 $\mathrm{km\,s^{-1}}$. We therefore adopt a representative value of $\sigma_{\rm loc}=1000\,\mathrm{km\,s^{-1}}$.

To explore possible non-virial motions, we further introduce a radial velocity component. In general, modeling genuine inflow or outflow motions requires detailed knowledge of the acceleration mechanism and initial conditions, e.g., those associated with radiation pressure or magnetic pressure \citep{Blandford1982b, Murray1995}. Such a treatment is difficult to prescribe from first principles and is beyond the scope of this work. We therefore adopt a phenomenological prescription in which the radial velocity of each cloud is set equal in magnitude to the tangential velocity of the corresponding Keplerian orbit\footnote{Here the phenomenological BLR model is treated in the Newtonian approximation, in contrast to the relativistic treatment adopted for the elliptical-disk model.}. This prescription produces larger radial velocities in the inner BLR and smaller radial velocities in the outer BLR. Although simplified, this behavior is qualitatively consistent with two plausible physical pictures: an accelerated inflow as gas spirals inward toward the black hole, and a decelerating outflow after gas is launched from the inner region by magnetic pressure. Moreover, introducing such radial motions affects only the overall width and the relative strength of the line core and wings, but does not break the symmetry of the velocity-resolved BDs.

\subsection{Construction of Synthetic Observational Signatures}\label{sec:2.3}

To construct the spatial distributions of the \ha\ and \hb\ emissivities across the BLR, we divide the elliptical disk described in Section~\ref{sec:2.2} into an $N_r \times N_\phi$ grid in polar coordinates. For each grid cell, the local \ha\ and \hb\ emissivities are assigned by interpolating the \textsc{Cloudy} emissivity tables obtained in Section~\ref{sec:2.1}. This procedure yields emissivity maps for \ha\ and \hb. By combining these emissivity maps with the velocity field and the light-travel-time delay of the BLR gas, we then construct the synthetic broad-line profiles, velocity-resolved BDs, and velocity-delay maps.

The emergent broad-line profile is the emissivity-weighted distribution of the BLR gas in velocity space. Therefore, the synthetic line profile can be obtained by integrating the emissivity map over the disk while projecting each grid cell onto its corresponding line-of-sight velocity:
\begin{equation} \label{eq:6}
    F(v_{\rm los}) = \sum_{j=1}^{N_r} \sum_{k=1}^{N_\phi} \epsilon_{jk}\,\Delta A_{jk}\,G(v_{\rm los}-v_{{\rm los},jk},\,\sigma_{\rm loc}),
\end{equation}
where $\epsilon_{jk}$ is the local emissivity in the grid cell $(j,k)$, $\Delta A_{jk}$ is the area of the cell, and $v_{{\rm los},jk}$ is its line-of-sight velocity. The Gaussian function $G(v_{\rm los}-v_{{\rm los},jk}, \sigma_{\rm loc})$ is a local broadening kernel. It describes how the emission from the grid cell is distributed in velocity space, accounting for its bulk line-of-sight velocity and local broadening effects. The velocity-resolved BDs are then calculated as the ratios between the synthetic \ha\ and \hb\ profiles in individual velocity bins.

In addition to the line-of-sight velocity, each grid cell is assigned a light-travel-time delay according to its location within the BLR. For a grid cell located at $(r_{jk}, \phi_{jk})$, the time delay is given by 
\begin{equation}\label{eq:7}
    \tau(r_{jk}, \phi_{jk}) = \frac{r_{jk}}{c}(1 - \sin i \cos \phi_{jk}).
\end{equation}
With this delay field, the characteristic BLR lag is obtained as the emissivity-weighted mean of the delays over the entire disk,
\begin{equation}\label{eq:8}
    \tau = \frac{\sum_{j=1}^{N_r} \sum_{k=1}^{N_\phi} \epsilon_{jk}\,\Delta A_{jk}\,\tau(r_{jk},\phi_{jk})}
    {\sum_{j=1}^{N_r} \sum_{k=1}^{N_\phi} \epsilon_{jk}\,\Delta A_{jk}}.
\end{equation}
It should be noted that RM measures the response of the line emission to continuum variations, and thus the observed lags correspond to a responsivity-weighted mean delay rather than an emissivity-weighted one. While the responsivity distribution is broadly similar to that of the emissivity \citep{Korista2004, Goad2014}, the emissivity-weighted lag adopted here should be regarded as a simplified approximation.

Beyond the characteristic lag, the distribution of the emissivity weights in two-dimensional velocity-delay space, also known as the velocity-delay map, encodes far more detailed information about the geometry and kinematics of the BLR. In RM studies, this map is commonly described by the transfer function $\Psi(v_{\rm los},\tau)$, which represents the response of the emission line at a given line-of-sight velocity and lag to variations in the driving continuum \citep[e.g.,][]{Horne2004}. 
The observed emission-line profile $F(v_{\rm los},t)$ is related to the continuum light curve $C(t)$ through
\begin{equation}\label{eq:9}
    F(v_{\rm los},t) = \int \Psi(v_{\rm los},\tau)\, C(t-\tau)\, d\tau.
\end{equation}

\begin{figure*}[t]
    \centering
    \includegraphics[width=\textwidth]{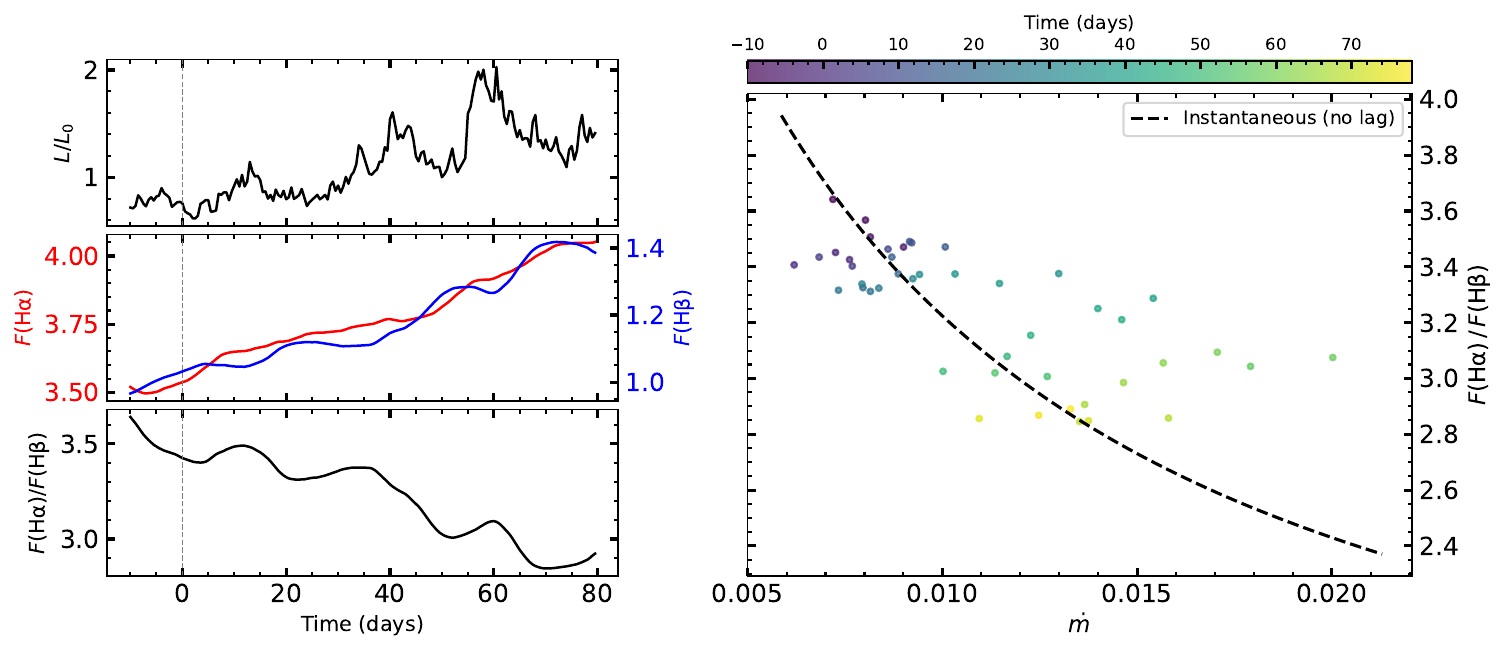}
    \caption{\textit{Left:} simulated AGN continuum variability (DRW, top), the resulting H$\alpha$ (red) and H$\beta$ (blue) flux light curves (middle), and the time-dependent BD (bottom). The BLR geometry is fixed and only the ionizing photon rate $Q({\rm H})$ varies with the continuum. \textit{Right:} BD versus accretion rate $\dot{m}$. The dashed curve shows the instantaneous (no-lag) relation, while the colored points include the reverberation time delay, color-coded by time. }
    \label{fig:lightcurve}
\end{figure*}

For an emissivity-weighted treatment, $\Psi(v_{\rm los},\tau)$ can be interpreted as the relative flux contribution from gas with line-of-sight velocity $v_{\rm los}$ and delay $\tau$. In our framework, $\Psi(v_{\rm los},\tau)$ can be constructed by stacking the emissivity-weighted contribution of each grid cell onto the $(v_{\rm los},\tau)$ plane:
\begin{equation}\label{eq:10}
\begin{split}
    \Psi(v_{\rm los},\tau) = \sum_{j=1}^{N_r} \sum_{k=1}^{N_\phi} \epsilon_{jk}\,\Delta A_{jk}\,G(v_{\rm los}-v_{{\rm los},jk},\,\sigma_{\rm loc})  \\
    \times \delta[\tau - \tau(r_{jk},\phi_{jk})],
\end{split}
\end{equation}
where the Dirac delta function $\delta$ assigns the contribution from each grid cell to its corresponding light-travel-time delay. Integrating $\Psi(v_{\rm los},\tau)$ over $\tau$ recovers the synthetic line profile, while its emissivity-weighted mean delay in each velocity bin,
\begin{equation}\label{eq:11}
    \tau(v_{\rm los}) = \frac{\int \tau\,\Psi(v_{\rm los},\tau)\,d\tau}{\int \Psi(v_{\rm los},\tau)\,d\tau},
\end{equation}
yields the one-dimensional velocity-resolved lags.

\section{Results} \label{sec:3}
In this section, we use a set of representative model parameters to investigate the observational signatures of the BD and the reverberation response of the BLR. Motivated by the observational features reported in \PaperI, we first examine how the integrated BD of the BLR varies with the incident ionizing continuum. We then construct velocity-delay maps and velocity-resolved BDs for virialized circular and elliptical disks, as well as for a circular disk with additional radial outflow motion, in order to assess the effects of disk geometry and kinematics. Finally, we compare the model predictions with observed emission-line profiles, velocity-resolved BDs, and velocity-resolved lags as a proof of concept.

\subsection{Relation between the Integrated BD and the Incident Continuum}\label{sec:3.1}

Some previous studies suggest that the intrinsic BD of the BLR is $\sim$3.0--3.2, with only a narrow range of variation. Thus, the BD has been widely used as a diagnostic of dust extinction in many AGN studies, with deviations from the canonical value usually attributed to reddening along the line of sight. However, in \PaperI, we found that the integrated BD can vary substantially on timescales as short as weeks to a few months, and that these variations are strongly anti-correlated with the incident ionizing continuum. Such behavior cannot be explained by variable extinction, but arises naturally in a photoionization framework \citep{Wu2023}, thereby providing a useful observational test for our photoionization-based BLR model.

\begin{figure*}[t]
    \centering
    \includegraphics[width=\textwidth]{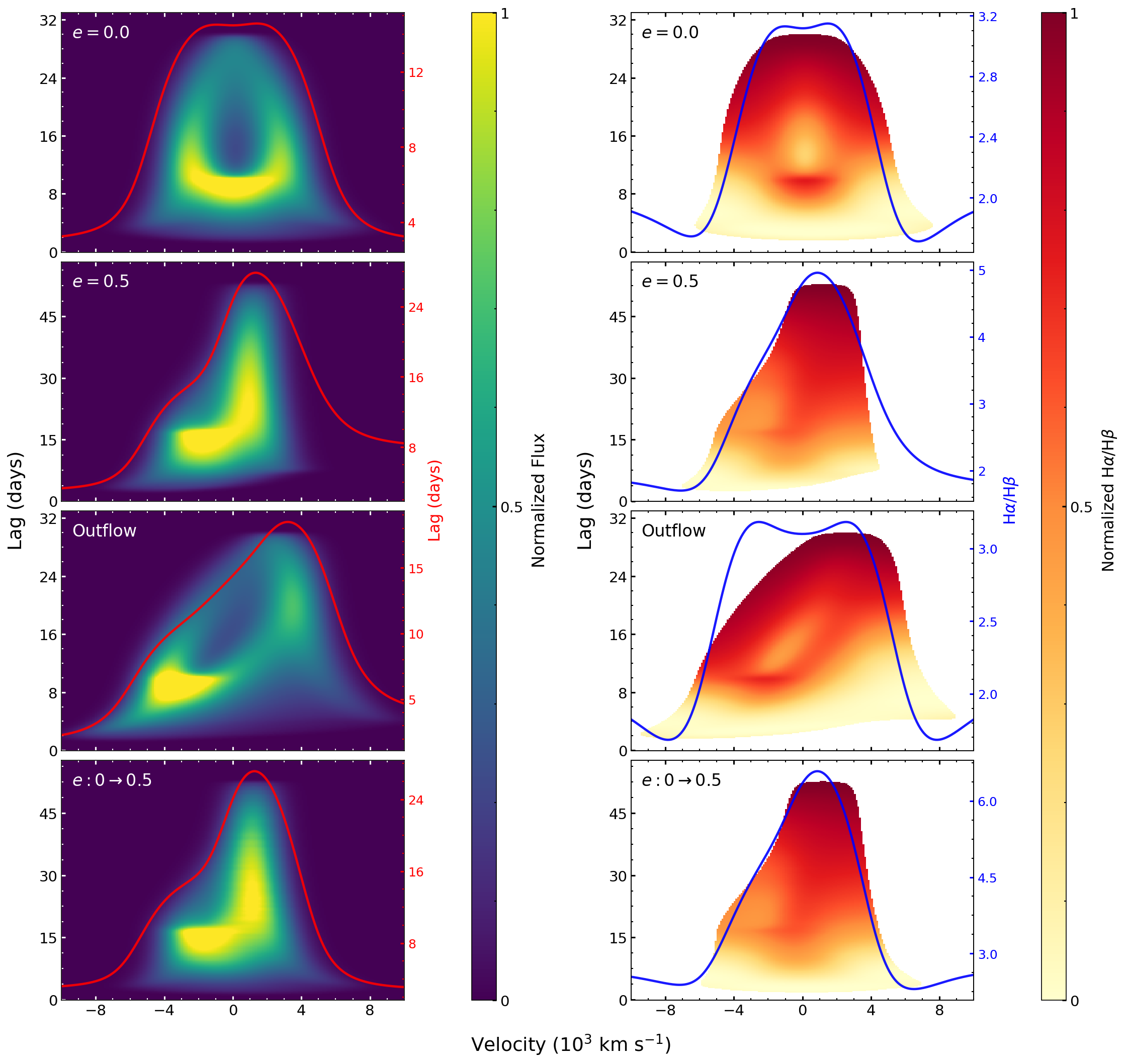}
    \caption{Simulation results for different disk configurations. From top to bottom, the panels show a circular disk ($e=0$), an eccentric disk ($e=0.5$), a circular disk with an additional radial outflow component ($v_r=v_\phi$), and a disk whose eccentricity gradually increases from $e=0$ at the inner edge to $e=0.5$ at the outer edge. \textit{Left column:} the H$\alpha$ transfer function $\Psi(v,\tau)$, with the intensity-weighted mean lag $\langle\tau\rangle(v)$ overplotted in red (right-hand axis). \textit{Right column:} the broad BD in the velocity--lag plane, with its velocity-resolved profile overplotted in blue (right-hand axis). 
    For clarity, the color scale in each panel is independently normalized to its own maximum. 
    Regions with \ha\ flux below 5\% of the corresponding maximum are masked to better display the detailed structure of the main BLR emission, since these low-flux regions contribute negligibly but can affect the displayed \ha/\hb\ color scale.}
    \label{fig:models}
\end{figure*}

To examine this effect, we calculate the integrated BD under different levels of the illuminating continuum. In this experiment, we adopt an elliptical disk model with an eccentricity of $e = 0.5$, extending from an inner pericenter $r_{\rm pi} = 500\,R_{\rm g}$ to an outer pericenter $r_{\rm po} = 3500\,R_{\rm g}$, where $R_{\rm g}$ is the gravitational radius. The local Gaussian broadening is set to $\sigma_{\rm loc} = 1000$ km s$^{-1}$, the disk inclination is $i = 30^\circ$, and the major-axis orientation is $\phi_0 = \pi/3$. Throughout the simulation, we neglect any evolution in the SED shape and allow only its normalization to vary. We adopt a black hole mass $\log (M_\bullet/M_\odot) = 8$ and a fiducial dimensionless accretion rate $\dot{m} = 0.01$, corresponding to an ionizing photon rate $\log Q({\rm H}) = 53.7$. With this set of parameters, the model gives an integrated BD of 3.22, close to the canonical value.

Since type 1 AGNs commonly exhibit optical flux variations by a factor of 2--3 on timescales of months to years, and such variability can often be described by a damped random walk (DRW) process, we generate a stochastic continuum light curve using a DRW model \citep[e.g.,][]{Macleod2010, Lu2019c}. We adopt the DRW variability amplitude and damping timescale to be $\sigma_{\rm DRW} = 0.4$ dex and $\tau_{\rm DRW} = 25$ days, respectively, comparable to those of NGC 3516 \citep{Feng2021b}. The simulated continuum light curve is shown in the upper-left panel of Figure~\ref{fig:lightcurve}. The resulting continuum variation spans a factor of $\sim$3, corresponding to an effective accretion rate range of  $\dot{m} \approx$ 0.006--0.02. Over this range, the integrated BD varies from $\sim$2.4 to $\sim$3.9. As shown in the right panel of Figure~\ref{fig:lightcurve}, the integrated BD follows a clear anti-correlation with the incident ionizing continuum, in good agreement with the observational trend reported in \PaperI (the green points in their Figure 1). This trend arises because changes in $Q({\rm H})$ affect the emissivities of \ha\ and \hb\ differently, causing the BD to evolve with the continuum state \citep{Wu2023}.

In the above calculation, the BD at each epoch is computed from the instantaneous incident continuum for all gas clouds within the BLR. This is an idealized treatment, because in a real BLR each gas cloud responds to the continuum variations with a time delay that depends on its location. In \PaperI, we found that the scatter in the observed BD--continuum relation is significantly reduced after correcting for the reverberation time delay using a DRW-based light-curve reconstruction, suggesting that the finite response time of the BLR can introduce additional scatter into the observed relation. To quantify this effect, we further calculate reverberation-delayed \ha\ and \hb\ light curves. As shown in Figure~\ref{fig:lightcurve}, these delayed emission-line light curves are smoother than the driving continuum. The reverberation effect also smooths the derived BD variations and introduces substantial scatter into the BD--continuum relation, consistent with observations. This demonstrates that reverberation effects can indeed blur the underlying anti-correlation between the integrated BD and the incident continuum.

\subsection{Geometric and Kinematic Effects on Velocity-Dependent Signatures}\label{sec:3.2}

Figure~\ref{fig:models} presents the two-dimensional velocity-delay maps and the distribution of the BD in the velocity--lag plane for different geometric and kinematic configurations, together with the corresponding one-dimensional velocity-resolved lags and velocity-resolved BDs. We adopt model parameters similar to those used in Section~\ref{sec:3.1}: an inner pericenter of $r_{\rm pi} = 500\,R_{\rm g}$, an outer pericenter of $r_{\rm po} = 3500\,R_{\rm g}$, a local Gaussian broadening of $\sigma_{\rm loc} = 1000$ km s$^{-1}$, a black hole mass of $\log (M_\bullet/M_\odot) = 8$, a dimensionless accretion rate of $\dot{m} = 0.01$, a major-axis orientation of $\phi_0 = \pi/3$, and an inclination of $i = 30^\circ$. We consider three representative cases: a virialized circular disk ($e = 0$), a virialized elliptical disk ($e = 0.5$), and a circular disk with an additional outflow component. The first two cases are used to examine the effect of eccentricity on the velocity-dependent lags and BDs, whereas the third case is used to test the impact of non-virial motion.

As shown in Figure~\ref{fig:models}, the velocity-delay map of the virialized circular disk exhibits a symmetric bell-shaped pattern, consistent with the prediction of \citet{Horne2004}. The corresponding BD distribution in the velocity--lag plane, as well as its one-dimensional velocity-resolved profile, is also essentially symmetric with respect to velocity. For the elliptical disk, both the velocity-delay map and the velocity-resolved lags are clearly asymmetric. Similarly, the BD distribution in the velocity--lag plane and its velocity-resolved profile become asymmetric. For the non-virial circular disk, the velocity-resolved lags resemble those of the elliptical disk, whereas the velocity-delay map differs in detail while remaining asymmetric. The BD distribution in the velocity--lag plane is likewise asymmetric, reflecting the underlying asymmetric delay distribution. However, after integration over lag, the one-dimensional velocity-resolved BD remains symmetric.

\begin{figure*}[t]
    \centering
    \includegraphics[width=\textwidth]{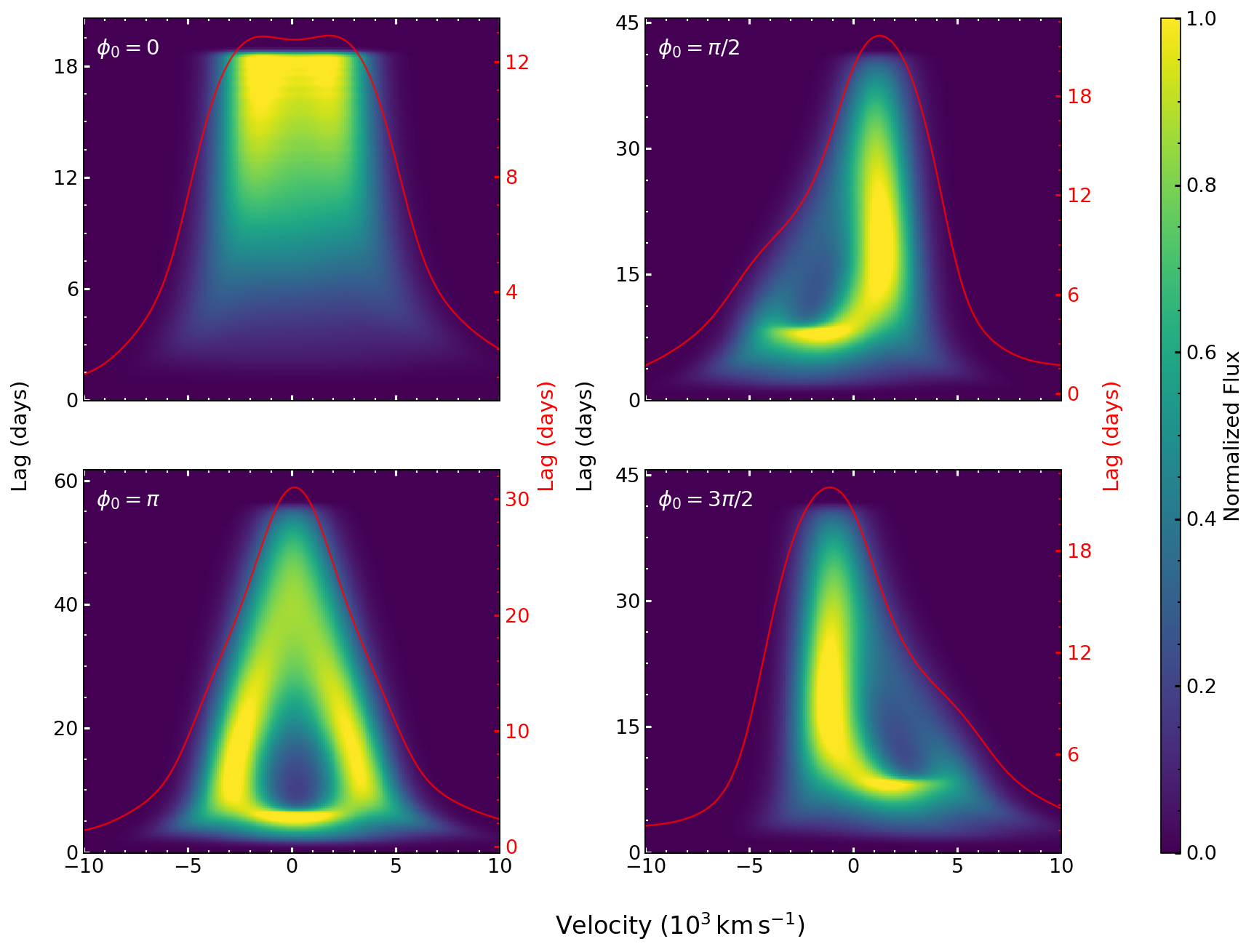}
    \caption{Effect of major-axis orientation $\phi_0$ on the \ha\, transfer function for a fixed eccentricity $e=0.5$ and inclination $i=30^\circ$. The red curve shows the intensity-weighted mean lag $\langle\tau\rangle(v)$. }
    \label{fig:3}
\end{figure*}

These results reveal a key diagnostic distinction. The velocity-resolved lags are subject to the degeneracy between geometry and kinematics: both an eccentric geometry and a radial outflow can produce similarly asymmetric lag profiles. The velocity-resolved BDs, in contrast, are more sensitive to the symmetry of the BLR itself. Combining these two observables therefore offers a promising way to isolate the kinematic contribution in complex BLR configurations.

\subsection{Orientation Effects on Velocity-Dependent RM Signatures}\label{sec:3.3}

The major-axis orientation of the elliptical disk plays an important role in shaping the asymmetric velocity-dependent response. To illustrate this effect, we generate velocity-delay maps and velocity-resolved lags for the virialized elliptical disk using the same parameters as in Section~\ref{sec:3.1}, varying only $\phi_0$.

As shown in Figure~\ref{fig:3}, changing $\phi_0$ alone can produce a wide range of response patterns, from nearly symmetric structures to asymmetric patterns in which either the red wing leads the blue wing or vice versa. In canonical RM studies, such signatures are often interpreted as evidence for different BLR kinematics, such as virialized motion, inflow, or outflow \citep[e.g.,][]{Denney2009, Wang2025_SAMP_V}. Our results demonstrate that a purely virialized elliptical disk can mimic all of these patterns. When combined with variations in eccentricity, inclination, and a finite BLR thickness, an even more diverse set of patterns emerges. This highlights a fundamental limitation of RM alone in constraining the geometry and kinematics of the BLR.

Beyond reshaping the velocity-dependent RM signatures, the orientation also affects the overall delay scale of the BLR. For example, the maximum lag is only $\sim 20$ days at $\phi_0 = 0$, whereas it can reach $\sim 60$ days at $\phi_0 = \pi$ for otherwise identical parameters. Such a projection-dependent variation in the measured lag can consequently introduce uncertainties into black hole mass estimates. Indeed, \citet{Wu2026} find through numerical simulations that the combined effects of multiple parameters can lead to a scatter of more than one order of magnitude in RM-measured lags.

\begin{figure*}[t]
    \centering
    \includegraphics[width=\textwidth]{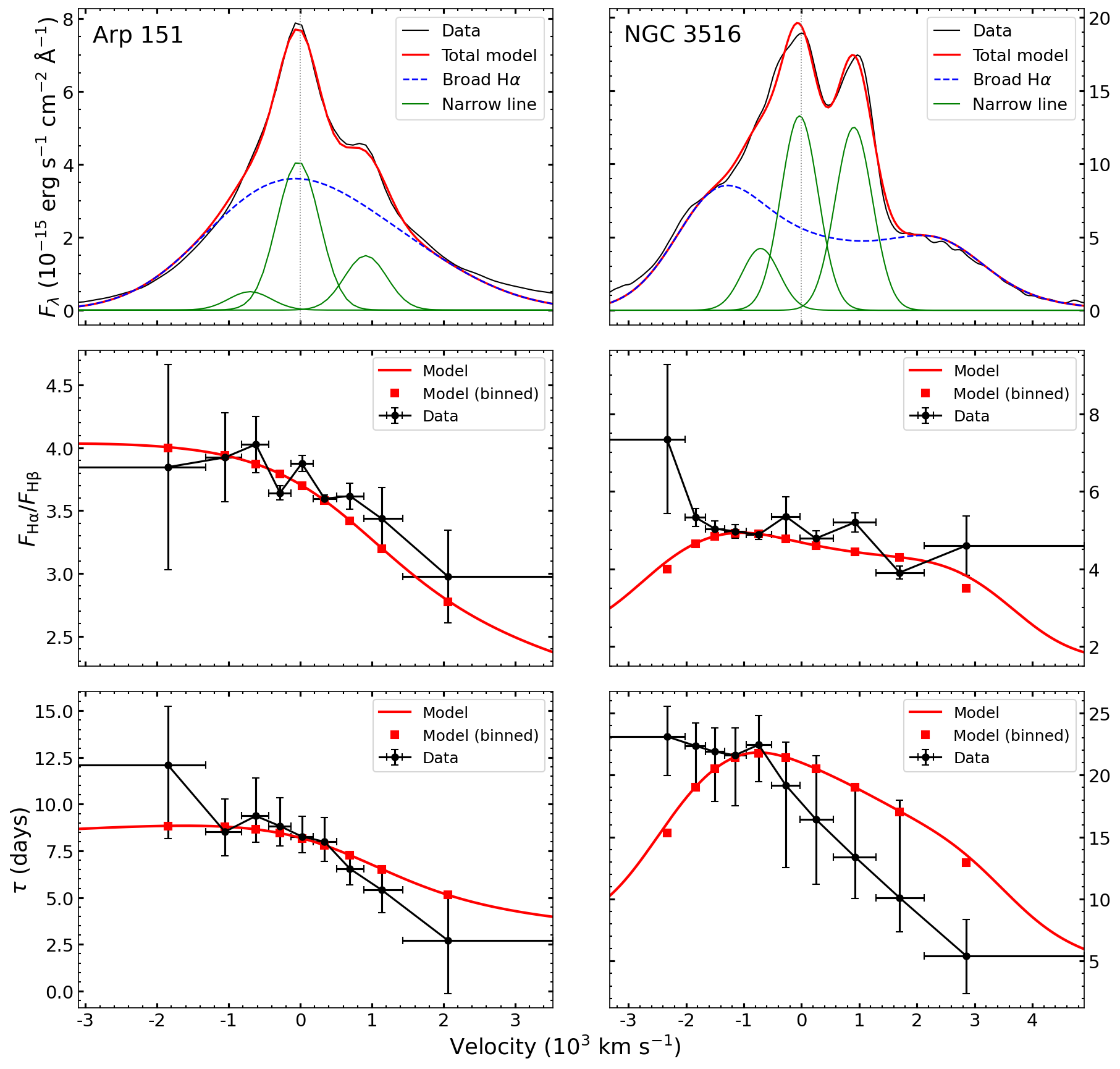}
    \caption{\textit{Top:} Decomposition of the \ha\ line profiles, including the broad components modeled with the elliptical disk (blue dashed), each narrow-line component modeled with a single Gaussian (green), and the total model (red). \textit{Middle:} Comparison between the observed velocity-resolved BDs and those predicted from the best-fitting line-profile parameters. \textit{Bottom:} Comparison between the observed \ha\ velocity-resolved lags and those predicted from the best-fitting line-profile parameters. The adopted model parameters are: Arp~151 --- $\log (M_\bullet/M_\odot) = 6.8$, $e = 0.5$, $\phi_0 = 1.3\pi$, $i = 30^\circ$, $\sigma_{\rm loc} = 1000$~km~s$^{-1}$, $r_{\rm pi} = 6000\,R_{\rm g}$, $r_{\rm po} = 13000\,R_{\rm g}$, $\log Q({\rm H}) = 53.0$; NGC~3516 --- $\log (M_\bullet/M_\odot) = 7.7$, $e = 0.28$, $\phi_0 = 1.3\pi$, $i = 38^\circ$, $\sigma_{\rm loc} = 600$~km~s$^{-1}$, $r_{\rm pi} = 1400\,R_{\rm g}$, $r_{\rm po} = 7400\,R_{\rm g}$, $\log Q({\rm H}) = 53.5$.}
    \label{fig:4}
\end{figure*}

\subsection{A Proof-of-Concept Comparison with Observations}\label{sec:3.4}

To assess whether our elliptical-disk model can account for real observations, we compare it with two representative AGNs. The first (Arp~151) exhibits double-peaked broad emission-line profiles, as expected for an elliptical-disk BLR. The second (NGC~3516) shows only single-peaked profiles, but both its velocity-resolved lags and velocity-resolved BDs display asymmetric patterns. A simultaneous fit to all of these observational features requires sophisticated photoionization calculations and flexible BLR modeling, which is beyond the scope of this work. Here, we adopt several simplifications intended only as an illustrative proof of concept.

We first construct the continuum-subtracted emission-line profiles based on the multi-Gaussian or Gauss-Hermite broad-line fits presented in \PaperI, and then fit the broad \ha\ and \hb\ line profiles using the photoionization-based elliptical-disk model described in Sections \ref{sec:2.1} and \ref{sec:2.2}. The narrow-line components are modeled with Gaussian profiles, with their velocity shifts and widths tied together. In the BLR modeling, we adopt black hole masses of $\log (M_\bullet/M_\odot) = 6.8$ for Arp 151 and $\log (M_\bullet/M_\odot) = 7.7$ for NGC 3516. The measured accretion rates are $\dot{m} \sim 0.05$ for Arp 151 \citep{Pancoast2014, Valenti2015} and $\dot{m} \sim 0.001$--$0.01$ for NGC 3516 \citep{Shapovalova2019, Shin2021, Mehdipour2022}.

In the \textsc{Cloudy} calculations, we adopt the approximate incident continuum SEDs for simplicity. Specifically, the SEDs for Arp 151 and NGC 3516 are computed using $(\log (M_\bullet/M_\odot), \dot{m}) = (7, 0.1)$ and $(8, 0.01)$, respectively. Following Section \ref{sec:3.1}, the total ionizing photon rate $Q({\rm H})$ is treated as a free parameter independent of the SED normalization, allowing the model to adjust the overall ionization level of the BLR. Based on the best-fit BLR parameters, we further generate the corresponding velocity-resolved BDs and velocity-resolved lags.

Figure~\ref{fig:4} shows the fitting results for the \ha\ line profiles, together with the corresponding velocity-resolved lags and velocity-resolved BDs predicted from the best-fitting line-profile parameters, which are compared with the observational results from \PaperI. Despite these simplifications, the elliptical-disk model can broadly reproduce these observational features simultaneously, suggesting that the geometric and kinematic framework developed in the preceding sections provides a plausible explanation for the observed signatures.

\section{Discussion}\label{sec:4}
\subsection{Eccentricity Gradients in the BLR} \label{sec:4.1}

In this work, we adopt an elliptical-disk model to describe an eccentric BLR. In our fiducial implementation, we ignore the finite thickness of the BLR and assume a radius-independent eccentricity. These simplifications allow us to isolate the basic effects of BLR eccentricity on the velocity-resolved lags and BDs. However, the actual eccentric structure of the BLR may be more complicated.

The physical origin of an eccentric BLR remains unclear. Several possible mechanisms have been proposed, such as gravitational perturbations from a binary SMBH \citep{Shen2010}, tidal disruption of clouds from the dusty torus \citep{Wang2017}, and failed disk winds or other processes that induce a non-axisymmetric perturbation in the BLR \citep{Czerny2017}. These scenarios do not necessarily produce a disk with the same eccentricity at all radii. Moreover, even if all BLR orbits initially have the same eccentricity, differential apsidal precession can naturally generate an eccentricity gradient across the BLR. Using Equation~(23) of \citet{Eracleous1995}, we estimate that, for the black hole mass adopted in this work, $\log (M_\bullet/M_\odot) = 8$, the precession timescale increases from $\sim 177$ yr at $r_{\rm pi} = 500\,R_{\rm g}$ to $\sim 2.3\times10^4$ yr at $r_{\rm po} = 3500\,R_{\rm g}$. This radial dependence implies that the inner BLR may circularize much faster than the outer BLR. Therefore, a uniform-eccentricity disk is unlikely to represent a fully realistic eccentric BLR over long timescales.

Motivated by this consideration, we perform an additional test with a radially varying eccentricity profile. In this model, the inner BLR is assumed to be circular, with $e=0$, while the outer BLR reaches $e=0.5$. For simplicity, the eccentricity increases linearly with radius. The results are shown in the bottom row of Figure~\ref{fig:models}. Although the detailed amplitudes and shapes of the velocity-resolved lags and velocity-resolved BDs are modified, their asymmetric patterns remain visible. This result suggests that the asymmetric signatures identified in our fiducial model are robust against the inclusion of a radial eccentricity gradient.

\subsection{Time-delay Effects on Velocity-resolved BDs}\label{sec:4.2}

\begin{figure}[t]
    \centering
    \includegraphics[width=0.47\textwidth]{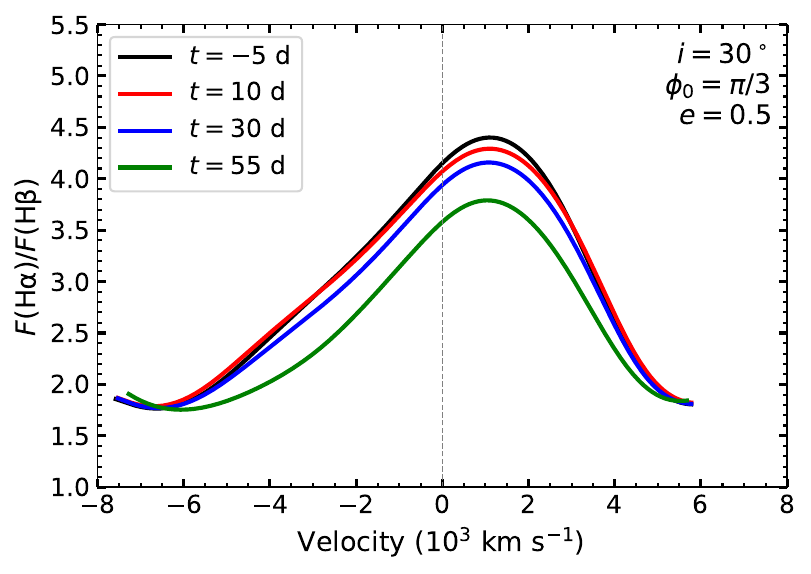}
    \caption{The evolution of the velocity-resolved BD profile.}
    \label{fig:profile_evo}
\end{figure}

As shown in Section~\ref{sec:3.1}, time delays introduce significant scatter in the relation between the integrated BD and the incident ionizing continuum. We therefore further examine whether such delays can also affect the shape of the velocity-resolved BDs. Following the same method as in Section~\ref{sec:3.1}, we calculate the velocity-resolved BDs at different delays.

The results are presented in Figure~\ref{fig:profile_evo}. The overall velocity-resolved BD profiles are nearly identical for different delays. This behavior is consistent with the measurements in Figure~5 of \PaperI, where the observed velocity-resolved BD profiles at different epochs show similar shapes. This weak dependence on delay can be understood as follows. In the BLR, the gas velocity is closely related to radius, whereas the incident ionizing flux decreases approximately as $r^{-2}$. The radial variation of the incident flux is therefore much stronger than the intrinsic luminosity fluctuations of the AGN over reverberation timescales. Consequently, the velocity-resolved BDs are primarily determined by the spatial distribution of the line-emitting gas rather than by short-term continuum variations.

In addition, the BD in a given velocity bin is a weighted average over gas from different positions within the BLR, which further reduces the imprint of time delays. A similar averaging effect also applies to the integrated BD. Although time delays introduce substantial scatter in the relation between the integrated BD and the incident ionizing continuum, they also reduce the variability amplitude of the integrated BD. Therefore, time-delay effects do not significantly change the overall shape of the velocity-resolved BD profiles.

\subsection{Implications}\label{sec:4.3}
Understanding the geometry and kinematics of the BLR is a fundamental goal of AGN studies. It is crucial not only for accurately measuring SMBH masses, but also for understanding the physical mechanisms operating in the innermost regions of AGNs, such as acceleration processes driven by radiation pressure or magnetic pressure \citep[e.g.,][]{Murray1995, Czerny2011, Netzer2013, Liu2024}. However, our results show that a virialized elliptical disk can reproduce a wide range of velocity-resolved lag patterns. This challenges the conventional kinematic interpretation of the diversity of RM signatures, in which symmetric lag patterns are attributed to virialized motion, whereas asymmetric lags are often interpreted as evidence for inflow or outflow. Therefore, our results indicate an intrinsic limitation of canonical RM analyses: velocity-resolved lags alone cannot uniquely distinguish between BLR geometry and kinematics. This degeneracy can hinder a robust reconstruction of the true spatial distribution and velocity field of the BLR. In this context, the IM method provides a critical additional constraint, because the velocity-resolved BD is sensitive to the symmetry of the BLR gas distribution and can therefore help break the degeneracy between non-axisymmetric geometry and non-virial motion.

The ability of an elliptical-disk model to simultaneously reproduce the observed line profiles, velocity-resolved lags, and velocity-resolved BDs lends strong support to the existence of an eccentric BLR. Such an eccentric geometry has severe practical consequences for RM-based black hole mass measurements. Because the time delay depends on the specific light-travel path, the emissivity-weighted lag of a non-axisymmetric BLR depends on how the elliptical gas distribution is projected along the line of sight. As demonstrated in Section \ref{sec:3.3}, varying only the orientation among a few representative values can already change the measured lag by a factor of $\sim3$. When combined with variations in eccentricity, inclination, $r_{\rm pi}/r_{\rm po}$, and other parameters, an even larger scatter is expected. At the same time, an eccentric BLR naturally produces complex broad emission-line profiles, further complicating the definition and measurement of the line width. \citet{Wu2026} demonstrate that an eccentric BLR can affect both the calibration of the virial factor and the interpretation of the empirical BLR radius--luminosity relation, and that these effects can together introduce more than an order-of-magnitude scatter into SMBH mass measurements.

These considerations underscore the important implications of BLR geometry for reliable black hole mass measurements in AGNs. Identifying and modeling non-axisymmetric BLR structures is therefore essential for improving the accuracy of SMBH mass measurements and for obtaining a more complete understanding of the physical conditions in AGN central regions. The joint RM+IM framework developed here offers a practical means to identify eccentric systems and thereby to flag, and ultimately correct, the geometry-induced biases that would otherwise remain hidden in conventional RM analyses.

\section{Summary}\label{sec:5}
We have constructed photoionization-based BLR models to quantitatively examine the diagnostic power of the joint RM+IM framework. By calculating the local \ha\ and \hb\ emissivities with \textsc{Cloudy} and coupling them with disk-like BLR models, we generated synthetic line profiles, velocity-delay maps, velocity-resolved lags, and velocity-resolved BDs. Our main results can be summarized as follows:

\begin{enumerate}
\item The integrated BD is expected to show large-amplitude variability and an anti-correlation with the incident ionizing continuum in our photoionization-based BLR model. Reverberation delays can introduce substantial scatter into the BD--continuum relation. This behavior is consistent with observations and demonstrates that the observed BD variability can be understood as a natural consequence of photoionization physics and finite BLR light-travel times.

\item Velocity-resolved RM signatures alone suffer from a strong degeneracy between geometry and kinematics. A virialized elliptical BLR can produce asymmetric velocity-delay maps and velocity-resolved lag profiles similar to those usually interpreted as signatures of inflow or outflow in conventional RM analyses, suggesting that velocity-resolved RM signatures alone cannot uniquely distinguish BLR geometry from kinematics.

\item The velocity-resolved BD is sensitive to the symmetry of the BLR gas distribution, and thus provides a complementary diagnostic of the BLR gas distribution. While both an eccentric disk and a circular disk with radial motion can produce asymmetric velocity-resolved lags, only the eccentric disk naturally yields correlated asymmetries in both velocity-resolved lags and BDs.

\item Proof-of-concept comparisons with one double-peaked and one single-peaked AGN show that an elliptical-disk BLR can simultaneously reproduce the observed line profiles, velocity-resolved lags, and velocity-resolved BDs, supporting the presence of eccentric BLR structures in AGNs. 
\end{enumerate}

These results provide a quantitative basis for IM and demonstrate that joint RM+IM analyses offer a promising way to break the BLR geometry-kinematics degeneracy. The presence of an eccentric BLR can significantly affect the measured lag and line width, and therefore has important implications for reverberation-based black hole mass measurements. More realistic BLR modeling and high-quality simultaneous monitoring of multiple emission lines will be crucial for applying this framework to larger AGN samples.

\begin{acknowledgments}
This work is supported by the National Key R\&D Program of China (No. 2021YFA1600404). J.C.W. and X.W.C. acknowledge support from the National Natural Science Foundation of China (NSFC, Grant Nos. 12533005, 12233007, 12347103). H.C.F. acknowledges support from NSFC (Grant Nos. 12573021, 12203096, 12633003), Yunnan Fundamental Research Projects (Grant No. 202601CK070011). S.S.L. acknowledges support from NSFC (Grant No. 12303022). We acknowledge support from the China Manned Space Project (CMS-CSST-2025-A07).
\end{acknowledgments}

\bibliography{sample631}
\bibliographystyle{aasjournal}

\end{document}